\documentclass[aps,prb,a4paper,twocolumn,showpacs,showkeys,floatfix, superscriptaddress]{revtex4-2}

\usepackage{graphicx}
\usepackage{color}
\usepackage{amsmath}
\usepackage{multirow}
\usepackage{hyperref}

\newcommand{\nabok}[2]{#1Cu$_7$(TeO$_4$)(SO$_4$)$_5$#2}
\newcommand{\cuso}{CuSO$_4 \cdot 5$H$_2$O}

\newcommand{\CsBr}{\nabok{Cs}{Br}}
\newcommand{\RbBr}{\nabok{Rb}{Br}}

\begin{document}

\title{Experimental observation of  decoupled spin subsystems in decorated square kagom\'{e} lattice magnets of the  nabokoite family}

\author{V.~N.~Glazkov}
\email{glazkov@kapitza.ras.ru}
\affiliation{ P.L.~Kapitza Institute for Physical Problems, RAS, Kosygina 2, Moscow 119334, Russia}

\author{Ya.~V.~Rebrov}
\affiliation{ P.L.~Kapitza Institute for Physical Problems, RAS, Kosygina 2, Moscow 119334, Russia}
\affiliation{National Research University  ``HSE'', 109028, Moscow, Russia}

\author{M.~A.~Dubovitskii}
\affiliation{ P.L.~Kapitza Institute for Physical Problems, RAS, Kosygina 2, Moscow 119334, Russia}
\affiliation{National Research University  ``HSE'', 109028, Moscow, Russia}

\author{M.~M.~Markina}
\affiliation{M.V.~Lomonosov Moscow State University, Moscow 119991, Russia}
\affiliation{National University of Science and Technology ``MISiS'', Moscow 119049, Russia}

\author{A.~F.~Murtazoev}
\affiliation{M.V.~Lomonosov Moscow State University, Moscow 119991, Russia}

\author{P.~S.~Berdonosov}
\affiliation{M.V.~Lomonosov Moscow State University, Moscow 119991, Russia}

\author{A.~N.~Vasiliev}
\affiliation{M.V.~Lomonosov Moscow State University, Moscow 119991, Russia}
\affiliation{National University of Science and Technology ``MISiS'', Moscow 119049, Russia}

\begin{abstract}
The square kagom\'{e} lattice (SKL) offers a model platform for investigating geometric frustration in 2D systems. Nabokoite-family compounds \nabok{A}{X} (A=Na, K, Cs, Rb and X=Cl, Br) extend this physics to a 3D network, where 2D SKL layers are decorated by interlayer spins. Using electron paramagnetic resonance (EPR), we demonstrate a dramatic splitting of this complex exchange network into two virtually decoupled spin subsystems: absolute calibration of the electron paramagnetic resonance (EPR) absorption reveals that only a fraction of  all copper spins in nabokoites is EPR-active and this fraction of the spins orders at the N\'{e}el point. Comparison of the EPR absorption and static susceptibility indicates that contribution of the EPR-silent spin subsystem to total magnetic susceptibility decreases on cooling. This direct observation of coexisting magnetic order and possible spin-liquid dynamics within a single compound challenges conventional models of unified exchange networks in decorated frustrated lattices.
\end{abstract}

\date{\today}



\maketitle

\section{Introduction}
\begin{figure}[th]
\centering
\includegraphics[width=\columnwidth]{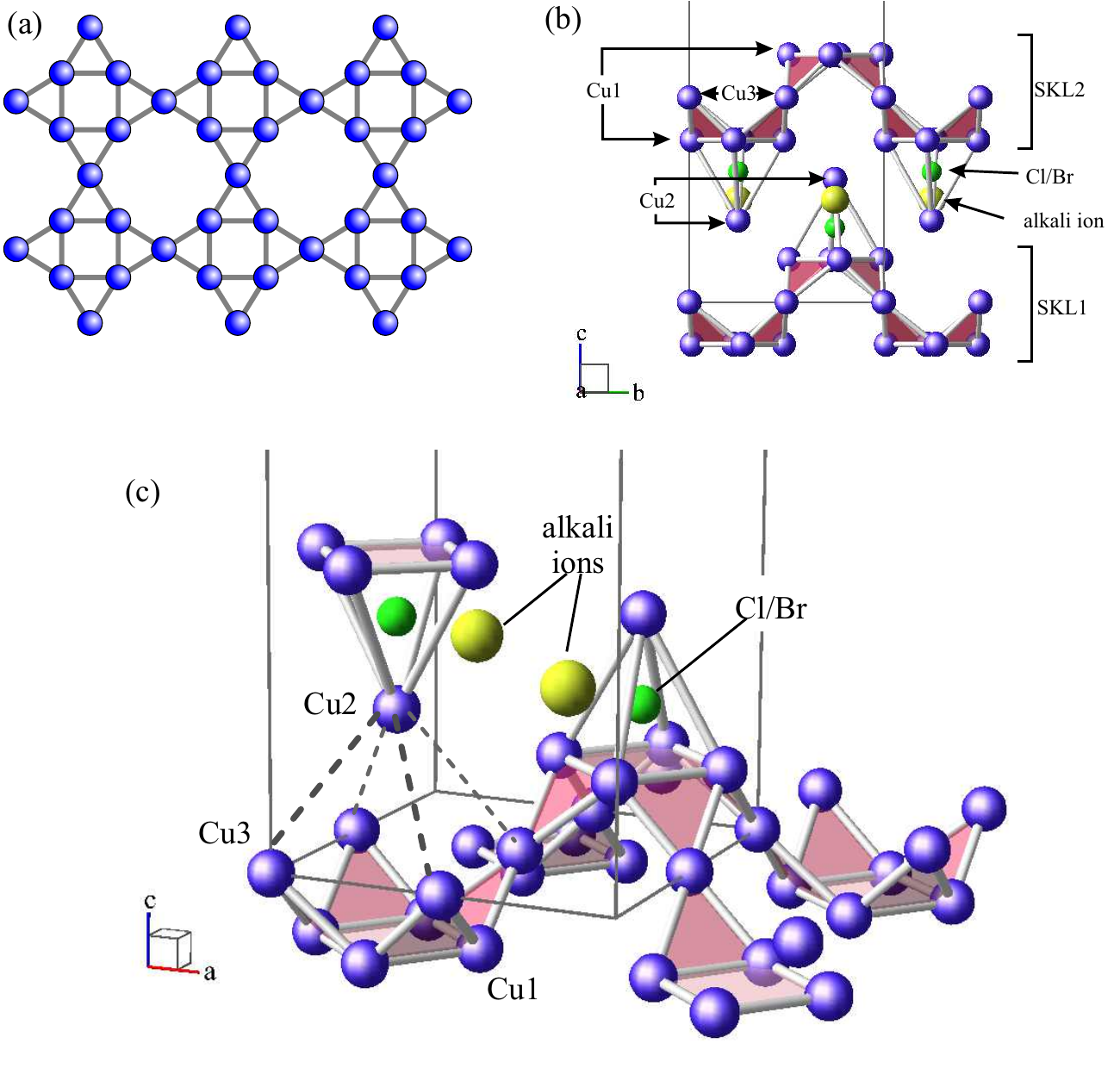}\\
  \caption{(color online) (a) Scheme of the ideal 2D kagom\'{e} lattice. (b) and (c) Fragments of the crystal structure of nabokoite family compounds \nabok{A}{X} with highlighted SKL layers. Only positions of copper, alkali and halogen ions are shown.}\label{fig:skl-and-struct}
\end{figure}

Square kagom\'{e} lattice (SKL), introduced by Siddharthan and Georges back in 2001 \cite{SidGeorge}, is an example of a strongly frustrated 2D lattice. Contrary to the usual kagom\'{e} lattice with hexagonal symmetry, square kagom\'{e} lattice features alternating square and octagon voids surrounded by triangles of spins with the net four-fold symmetry (Fig.~\ref{fig:skl-and-struct}-a). The original paper \cite{SidGeorge} suggested a gapped spectrum of triplet excitations in SKL magnet with low-lying singlet states originating from a mixture of dimer configurations along with possible hidden order in 2D SKL system formed by alternating distribution of electron density along sides of the squares and four-beam stars. Numerical simulations of the $S=1/2$ SKL antiferromagnet thermodynamics \cite{Richter-skl-gapped}
confirmed gapped magnetic excitations spectrum in the ``ideal'' (with the same exchange couplings along all bonds) system.

Extensions of this model  taking into account inequivalence of exchange couplings along the sides of triangle and the next-neighbor couplings \cite{morita-j123,Lugan,Gembe-noncomplanar} revealed high sensitivity of the SKL magnet ground state to the deviations from the ``ideal'' lattice. The variety of the achievable ground states includes, e.g.,  topological nematic ground state \cite{Lugan}, Ne\'{e}l states \cite{morita-j123} and complicated non-coplanar ordered states \cite{Gembe-noncomplanar}.

The quest for an experimental realization of the SKL model is still open, several minerals have been proposed to host SKL layers in their crystal structure \cite{fuji, yakubovich, bo-specheat,alisher}. Inelastic neutron scattering and muon rotation experiments reported in  Ref.~\cite{fuji} demonstrated presence of a gapless spin-liquid in one of these minerals, atlasovite KCu$_6$FeBiO$_4$(SO$_4$)$_5$Cl, where Fe was substituted by Al. This system remains disordered down to at least 50~mK. No direct observation of the gapped spin-liquid state predicted for the ``ideal'' SKL system has been reported so far.

Recently synthesized \cite{alisher} nabokoite family compounds \nabok{A}{X}{}, here ``A'' denotes alkali ion (Na, K, Cs or Rb) and ``X'' denotes chlorine or bromine, provide a particular example of the \emph{decorated} square kagom\'{e} lattice (Fig.~\ref{fig:skl-and-struct}-b,c). Six copper ions from the nabokoite formula unit occupy crystallographic positions Cu1 and Cu3  and form SKL layers, while the seventh ion occupies interlayer Cu2 position. These compounds were studied earlier by thermodynamic measurements (magnetization and specific heat), electron spin resonance and dielectric properties measurements \cite{alisher, markina,rebrov}. Recently, single-crystalline samples of \nabok{K}{Cl}{} were studied by using variety of techniques, including NMR \cite{china}.

In the present paper we report development of the electron paramagnetic resonance (EPR) and magnetization study of the nabokoite family magnets. By modifying  the experimental technique, including broad bi-polar field sweeps and accurate absolute calibration of the EPR absorption, we succeeded to observe broad (full linewidth about 1~T) paramagnetic resonance absorption line and to estimate the fraction of nabokoite spins contributing to the EPR absorption. The observed EPR absorption signal transforms to antiferromagnetic resonance absorption at the N\'{e}el point $T_\textrm{N}$, but its absolute intensity at $T\gg T_\textrm{N}$ corresponds to about one-half of the nabokoite sample total spin susceptibility. This experimental finding directly reveals separation of the spin subsystems in nabokoites: only a fraction of spins in nabokoite remains EPR-active, and the same fraction orders below the N\'{e}el point. Remaining spins contribute to the static magnetization, but are  invisible for electron spin resonance, most likely due to the even larger linewidth. The lacking EPR intensity decreases at low temperature providing indications  of the formation of the gapped state within the EPR-silent subsystem.

\section{Experimental details and samples}

\subsection{Samples}
Polycrystalline samples of nabokoite family compounds \nabok{A}{X}{} (A=K, Na, Rb, Cs and X=Cl, Br) were produced by solid state synthesis as described in Ref.~\cite{alisher}.  Samples with all possible combinations of alkali (A) and halogen (X) ions save for the Na/Br combination were successfully synthesized \cite{alisher,alisher-phd}. We will refer to these compounds as ``A/X nabokoites'' for short.

The crystal structures of the samples were checked by X-ray diffraction, all samples were found to be  isostructural with P4/ncc (D$_\textrm{4h}^8$) space group at room temperature. Both alkali and halogen ions are located outside of the SKL layer, so the main effect of chemical composition change is the stretching along the tetragonal $c$-axis. Indeed \cite{alisher,alisher-phd}, chlorine to bromine substitution increases lattice parameter $c$ by approximately 2\% and variation of alkali ion from light (Na or K) to heavy (Cs) increases lattice parameter $c$  by approximately 5\%, while change of the lattice parameter $a$ remains about or below 0.8\%.

Previous studies  \cite{markina,rebrov,china,nabok-afm} revealed dielectric anomalies in light-alkali-ion (Na, K) nabokoites hinting at possible minute structure modifications below 100~K. Heavy-alkali-ion (Rb,Cs) nabokoites do not demonstrate significant dielectric anomalies above the N\'{e}el point. For this reason we chose Rb/Br and Cs/Br nabokoite samples for wide temperature range quantitative  EPR study while other nabokoites samples were used only for qualitative observation of the paramagnetic absorption.

Magnetization and specific heat curves of \nabok{Rb}{Br}{} and \nabok{Cs}{Br}{} were measured to control samples quality and to check for the magnetic phase transitions. Magnetization, specific heat and electron spin resonance experiments confirm that Rb/Br nabokoite orders antiferromagnetically at $(5.44\pm0.10)$~K and Cs/Br nabokoite orders antiferromagnetically at $(5.25\pm0.10)$~K. Details of the low-temperature antiferromagnetic phase study will be reported separately \cite{nabok-afm}.

\subsection{Experimental techniques}
Magnetization curves $M(B,T)$ were  measured at Moscow State University using Quantum Design MPMS-7T SQUID magnetometer. Samples for the magnetization measurements were prepared by slightly pressing nabokoites powder within the plastic sample-holder container.

Electron spin resonance experiments were done at Kapitza Institute using home-made transmission type spectrometers of different types:  spectrometer with 1~T water-cooled magnet was used for the 9-14~GHz experiments at 77-300~K temperature range and spectrometer with 7~T helium-cooled superconducting magnet was used  for the 9-150~GHz experiments at 1.5-250~K temperatures. Frequency range 9-150~GHz was covered by two spectroscopic insets with cavities of different size and a variety of microwave generators.

Samples for EPR measurements were prepared by soaking 50-150~mg of nabokoite powder with ethanol diluted glue (BF-2 trademark,  which is a close analogue to GE varnish) within the thin-walled paper container (container diameter 5~mm, sample height 1-2~mm). Sample was placed on the wall of copper microwave cavity which was surrounded by a vacuum shell submerged in a liquid nitrogen (water-cooled magnet) or liquid helium (superconducting magnet) bath. Vacuum shell could be filled with a small amount of heat exchange helium gas for the experiments close to the bath temperature or  vacuumized for high-temperature measurements. For high temperature experiments, when the cavity temperature is well above the bath temperature, we additionally control temperature gradient over the cavity length to ensure temperature measurement conditions: at cavity temperature around 200~K and liquid helium  bath (4.2~K) the temperature gradient over the cavity length was found to be  within $\Delta T=3$~K.

In the case of nabokoites the  observed paramagnetic resonance line is quite broad: its linewidth (about 1~T) is comparable to the paramagnetic resonance field (which  is about 1~T at 30~GHz microwave frequency). This complicates accurate determination of the absorption line parameters  as the  zero-field absorption handicaps baseline determination. To overcome this issue and to increase reliability of data processing  superconducting magnet setup of the EPR spectrometer was modified to allow for bi-polar field sweeps. The  parameters of the background drift can be then reliably determined comparing signals at the extreme left ($-B_\textrm{max}$) and right ($+B_\textrm{max}$) parts of a bi-polar field sweep.

\subsection{Details of electron spin resonance absorption analysis}
Electron paramagnetic resonance is the resonance absorption of the microwave radiation when microwave quantum coincides with Zeeman splitting of spin sub-levels $\hbar \omega=g\mu_\textrm{B} B_\textrm{res}$. For the linearly polarized microwave field the absorption should appear symmetrically at $\pm B_\textrm{res}$. Position of the resonance absorption provides information on $g$-factor, EPR linewidth is related to the spin relaxation processes and EPR absorption integral intensity can be linked to the spin susceptibility of the system \cite{altkoz,a-blean}.

Our experimental setup does not use field modulation. We measure microwave power transmitted through the microwave cavity with the sample, for small absorption change of the transmitted signal is proportional to the sample absorption $\chi''$. Some technical factors (e.g., small de-tuning of microwave generator frequency from the cavity eigenfrequency) add weak contribution of the dispersion $\chi'$ to the observed signal. Assuming  Lorentzian shape absorption, microwave power transmitted through the cavity with the sample can be described by the following equation:

\begin{equation}\label{eqn:p-trans}
\begin{split}
P_\textrm{trans}=P_0 (1+k B)(1-\frac{A_1+D_1(B-B_\textrm{res})}{1+(B-B_\textrm{res})^2/\Delta^2}- \\
-\frac{A_2-D_2(B+B_\textrm{res})}{1+(B+B_\textrm{res})^2/\Delta^2})
\end{split}
\end{equation}

\noindent here $P_0$ is the microwave power transmitted through the cavity without resonance absorption, term $(1+k B)$ phenomenologically describes background drift of the transmitted signal, e.g., due to the slow drift of microwave generator parameters. Absorption amplitudes $A_{1,2}$ should be the same in the ideal case, however, they are highly sensitive to detuning of microwave generator away from the cavity eigenfrequency and we have found that the observed line-shape is better fitted assuming both amplitudes as free parameters. Dispersion parameters $D_{1,2}$ should be zero in the ideal case, they depend on microwave generator detuning and were considered as  free parameters in a lineshape fit procedure. EPR linewidth $\Delta$ and resonance field $B_\textrm{res}$ are the same for the positive and negative field absorption signals.  Integral intensity of the Lorentzian absorption line is $I_i=\pi A_i \Delta$.

Powdered samples require additional considerations as absorption signal is averaged over all powder particles orientations. This averaging usually produces characteristic  absorption signal spread in the field range determined by the extreme $g$-factor values with sharp left and right edges. In the case of nabokoites EPR absorption linewidth $\Delta$ is about 1~T and exceeds by far possible change of the resonance absorption field due to the $g$-factor anisotropy at our experimental conditions: $\delta B_\textrm{res}\simeq \frac{\delta g}{g}B_\textrm{res}\simeq 0.1B_\textrm{res}$ for typical Cu${}^{2+}$ ion $g$-factor anisotropy with $B_\textrm{res}\simeq 1$~T at microwave frequency $f\simeq 35$~GHz. This allows us to use single Lorentzian line fit \eqref{eqn:p-trans} to describe paramagnetic absorption in nabokoites. At the same time paramagnetic defects used for the absolute calibration of EPR absorption have a much narrower intrinsic resonance line yielding a conventional asymmetric ``powder absorption'' line-shape. Integral intensity of such an asymmetric signal was determined by numeric integration.

\section{Experimental results}
\subsection{Paramagnetic resonance absorption in nabokoite family magnets}
\begin{figure}[th]
\centering
  \includegraphics[width=\columnwidth]{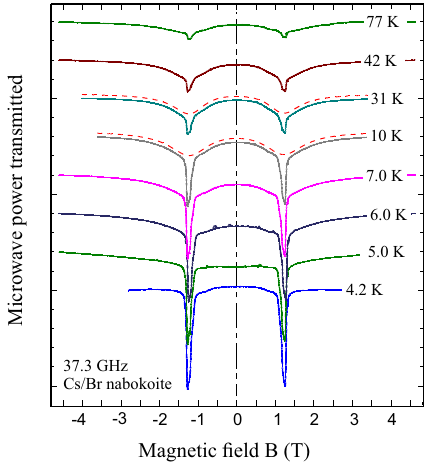}\\
  \caption{(color online) Paramagnetic resonance absorption spectra (bi-polar field sweep) for powder sample of \nabok{Cs}{Br}{} at different temperatures. Dashed curves show Lorentzian fits of 10 and 31~K data using Eqn.~\eqref{eqn:p-trans} (fits are shifted upwards for clarity). }\label{fig:PMspectra}
\end{figure}

\begin{table}[th]
\caption{Characteristic paramagnetic resonance linewidth $\Delta B$ (half-width at half-maximum) in different nabokoite family compounds \nabok{A}{X}. }\label{tab:width}
\begin{ruledtabular}
\begin{tabular}{ccc}
Compound (A/X)&$\Delta B$, T&Exp. conditions\\
\hline
Na/Cl&0.57&38.7~GHz, 20~K\\
K/Cl&0.51&32.8~GHz, 107~K\\
Rb/Cl&0.32&38.5~GHz, 25~K\\
Cs/Cl&0.71&99.6~GHz, 10~K\\
K/Br&0.57&28.4~GHz, 15~K\\
Rb/Br&0.60&38~GHz, 100~K\\
Cs/Br&0.68&38~GHz, 100~K\\
\end{tabular}
\end{ruledtabular}
\end{table}

Examples of magnetic resonance absorption spectra for Cs/Br nabokoite at different temperatures are shown at Fig.~\ref{fig:PMspectra}. Similar electron paramagnetic resonance  signals were observed for all \nabok{A}{X}{} nabokoites samples.  Above the N\'{e}el point (which is around 5.2~K for Cs/Br compound) resonance absorption spectrum consists of a broad Lorentzian line with the line width (half-width at half-maximum, HWHM) about 0.5-0.6~T and a narrow absorption signal with asymmetric line-shape typical for the powder averaged paramagnetic resonance absorption. At the N\'{e}el temperature this broad absorption line changes, and depending on the microwave frequency it shifts, splits or even disappears on further cooling. Details of magnetic resonance in the antiferromagnetic phase will be discussed elsewhere \cite{nabok-afm}. For the purposes of the present discussion it is important to note, that for all \nabok{A}{X}{} nabokoites samples the observed broad paramagnetic line transforms smoothly into the antiferromagnetic resonance signal below the N\'{e}el point which is a direct indication that the long-range order is formed within the spin subsystem giving rise to the observed paramagnetic absorption. The narrow component does not change its position or line-shape at the transition point, its intensity increases on cooling in  agreement with the Curie law \cite{markina}.

This allows us to ascribe the narrow powder-averaged absorption signal to some kind of free paramagnetic centers, most likely from some magnetic defects of the sample, totally decoupled from the main spin system of nabokoite.  Similar powder-averaged absorption was observed in our earlier research on K/Cl nabokoite \cite{markina}.  Boundaries of this powder-averaged signal correspond to the $g$-factor values  2.05 and 2.35 which are typical for copper ions.  Concentration of these centers is around 1-3\% per copper ion as will be determined in the next subsection. The exact microscopic origin of these paramagnetic centers (e.g., volume or surface defects)  is not important for the present discussion, regardless of it we can use the paramagnetic absorption of these centers to scale the absorption intensity related to the broad absorption component.

The broad EPR absorption signal evaded detection  in our previous research \cite{markina} because of its large linewidth and relatively small amplitude. Use of the bi-polar field sweeps from $-5$~T to $+5$~T make presence of this absorption component clear, its contribution can be well described by a Lorentzian fit \eqref{eqn:p-trans}. Linewidth (HWHM) of the broad EPR absorption line for different compounds vary from 0.3~T  to 0.7~T (see Table~\ref{tab:width}).

By comparing observed absorption spectra we have chosen Cs/Br and Rb/Br nabokoites for the detailed study of temperature dependence of EPR absorption. This choice was justified by the following considerations: (i) absence of dielectric anomalies in heavy-alkali ion nabokoites (strong dielectric absorption in the  vicinity of anomaly \cite{rebrov} highly complicates electron spin resonance measurements), and (ii) convenient relation between amplitudes of the absorption signals from the paramagnetic centers and from the decorated SKL subsystem in these samples (which is the key for the accurate absorption calibration as will be discussed below).

\subsection{Experimental procedure for the quantitative calibration of EPR absorption}
\begin{figure}[th]
\centering
  \includegraphics[width=\columnwidth]{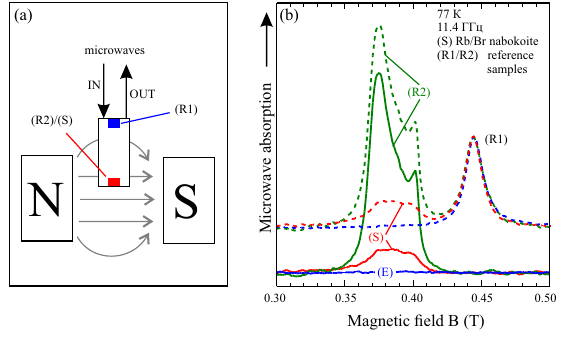}\\
  \caption{(color online) (a) Sketch of the cavity and sample positions at the calibration of paramagnetic centers absorption: (S) --- nabokoite sample placed on the lower cavity wall (uniform field of the magnet), (R1) --- reference single-crystalline sample of \cuso{} placed on the upper cavity wall (stray field of the magnet), (R2) --- reference powdered sample of \cuso{} replacing the sample. (b) Comparison of the scaled EPR absorption spectra in calibration experiments with Rb/Br nabokoite ($f=11.4$~GHz, $T=77$~K, $m_\textrm{S}=81.2$~mg, $m_\textrm{R2}=5.5$~mg): sample (S)+reference sample (R1); powdered reference sample (R2) + reference sample (R1); no sample on the lower wall (E)+reference sample (R1). Dashed curves ---  EPR absorption scaled to the same intensity of (R1) absorption signal, solid curves --- scaled EPR absorption curves with (R1) signal subtracted. }\label{fig:2samples}
\end{figure}

\begin{table}[th]
\caption{Concentration of the free paramagnetic centers in \nabok{A}{X} samples, per copper spin in decorated SKL sub-system of nabokoite. Na/Br nabokoite sample was not available. Reliable measurements in K/Br nabokoite sample at liquid nitrogen temperature were impossible because of closely located dielectric anomaly. }\label{tab:intdef}
\begin{ruledtabular}
\begin{tabular}{ccccc}
&Na&K&Rb&Cs\\
Cl&2.1\%&3.5\%&0.45\%&$<$0.1\%\\
Br&no sample&---&0.91\%&1.12\%\\
\end{tabular}
\end{ruledtabular}
\end{table}

The key experimental result of the present paper relies on the accurate \emph{quantitative} calibration of the EPR absorption in nabokoites.

Absorption of the  microwave radiation by the sample in a microwave cavity is given by the equation

\begin{equation}\label{eqn:abs}
\frac{\Delta P_\textrm{trans}}{P_0}\propto \eta \chi'' Q
\end{equation}

\noindent here   $\chi''(\omega,B)$ is the  imaginary part of high-frequency magnetic susceptibility (per unit volume), $Q$ is the unloaded cavity Q-factor and the filling factor $\eta$ is the ratio of microwave energy density integrated over the sample volume $V_\textrm{sample}$ to the microwave energy density integrated over the cavity volume (see, e.g., \cite{pool}). Integral intensity of the paramagnetic resonance absorption $\int \chi''(\omega, B) dB$ is proportional to the sample static susceptibility.

The main complication of the quantitative analysis of EPR absorption is the presence of a variety of hardly controllable experimental factors (e.g., cavity Q-factor or high-frequency magnetic field distribution within the cavity). We took advantage of the presence of free paramagnetic centers in the nabokoite samples, which are subject to the exactly the same experimental conditions and provide a distinct EPR response. Quantity of these free paramagnetic centers can be determined in a separate experiment via comparison with the paramagnetic standard. The experimental procedure used is described below in detail.

The calibration protocol includes following steps: (i) determination of the concentration of the free paramagnetic centers  using reference samples; (ii) scaling of the broad EPR absorption against the  free paramagnetic centers resonance absorption; (iii) calculation of the contribution of the EPR-active spins to the total static susceptibility of nabokoites assuming Curie law for the susceptibility of the free paramagnetic centers.

(I) As  a first step we performed the EPR experiments at 77~K (liquid N$_2$) temperature using the  low-frequency (11~GHz) spectrometer equipped with water-cooled 1~T magnet. The $g=2.00$ resonance field at $f=11$~GHz is 0.39~T, which is less than the typical linewidth of the broad EPR absorption component. Hence,  only the paramagnetic resonance signal from the free paramagnetic centers is observed. Our goal at this step is to compare intensity of the free paramagnetic centers absorption in the nabokoite sample (S)  against  the paramagnetic resonance absorption in the reference sample. We chose copper sulphate pentahydrate \cuso{} as a reference sample because of the convenient EPR linewidth and easy availability of these crystals.  Crystals of \cuso{} for the calibration experiments were re-crystallized from the water solution by slow water evaporation at ambient conditions. Curie-Weiss temperature for \cuso{} is about 1~K \cite{cuso4}, so at liquid nitrogen temperature it can be considered as an ideal paramagnet.

Scheme of the experiment is shown at Fig.~\ref{fig:2samples}-a. Microwave cavity with the length $L=40$~mm was placed between the poles of the magnet so that the lower wall of the cavity is within the uniform magnetic field area while the upper wall of the cavity is in the stray field area.  Small \cuso{} crystalline sample (reference sample (R1) at Fig.~\ref{fig:2samples}-a) with the mass of about 1.5-2.5~mg was placed at the upper wall of the cavity. The upper wall of the cavity is located in the magnet stray field and resonance conditions for this sample are fulfilled at the larger field at magnet center, which allows us to separate (R1) sample absorption. The reference sample (R1) on the upper wall of the cavity remains unchanged during the calibration experiment. The nabokoite powder sample (S) with the mass $m_\textrm{S}=81.2$~mg or the powdered \cuso{} sample (R2) with the mass $m_\textrm{R2}= 5.5$~mg were alternately placed on the lower wall of the cavity.  Possible cavity background was checked in an additional experiment with empty lower cavity wall (E). Samples (S) and (R2) are placed in the same position, large wavelength at 11~GHz ensures practically uniform microwave field at sample position. In this case the  filling factor $\eta$ in Eqn.~\eqref{eqn:abs} is proportional to the sample volume and integrated intensity of the absorption is proportional to the total static spin susceptibility of the sample.

Measured microwave absorption spectra (see Fig.~\ref{fig:2samples}-b) are scaled to the same integral intensity of the  reference sample (R1) absorption. The (R1) sample is exactly the same and is located at the same place in all three experiments ((R1)+(S);(R1)+(R2);(R1)+(E)). This scaling eliminates uncertainties related to the change of the cavity configuration at each sample mounting. Nabokoite powder sample (S) and powdered reference sample (R2) are subjects to the powder-averaging over the anisotropic $g$-factor values, which reduces possible $g$-factor averaging corrections.

We then  compare  integral intensity  of  the scaled  free paramagnetic centers absorption signal (S), $I_\textrm{S}$,  against the  intensity  of the scaled EPR absorption in reference sample (R2), $I_\textrm{R2}$, and  calculate concentration of the free paramagnetic centers $x$

\begin{equation}\label{eqn:defects}
x=\frac{1}{7}\frac{I_\textrm{S}}{I_\textrm{R2}} \frac{m_\textrm{R2}/\mu_\textrm{R2}}{m_\textrm{S}/\mu_\textrm{S}}
\end{equation}
\noindent here $\mu_\textrm{S, R2}$ are the corresponding molar masses, factor 1/7 corresponds to the calculation of the defects concentration per copper spin in nabokoite. Accuracy of this calibration is limited by the experimental noise and some remaining technical issues. We estimate effect of these uncertainties to be about 10-20\%.

This approach allowed us to determine concentration of the free paramagnetic centers in all studied nabokoite family compounds, see Table~\ref{tab:intdef}. It was found to be around several percents or less for all studied samples. Measurements with K/Br nabokoite sample were not possible because of the dielectric anomaly located close to the experiment temperature  \cite{nabok-afm}.  Results for the free paramagnetic centers concentration in  K/Cl nabokoite are close to the earlier estimate of Ref.~\cite{markina}.

(II) As a second step of the calibration protocol we measured EPR response of the same nabokoite sample at higher frequencies (around 35~GHz) using a EPR spectrometer with superconducting magnet and applying bi-polar field sweeps from $-5$~T to $+5$~T. As  the field sweep amplitude exceeds the linewidth of the broad component, we can resolve both EPR signal from nabokoite spins and from the free paramagnetic centers  (Fig.~\ref{fig:PMspectra}). Microwave wavelength at 35~GHz is comparable with the sample size, hence the  microwave field distribution within the nabokoite sample is unknown. However, as the free paramagnetic centers are equally distributed over the nabokoite powder sample, this distribution is \emph{exactly the same} for both spin systems. Hence, the ratio of the integral intensity of the broad EPR absorption from the nabokoite spins, $I_\textrm{nabok}$, to the integral intensity of the  powder-averaged EPR absorption from the free paramagnetic centers, $I_\textrm{free}$,  depends only on the ratio of static spin susceptibilities of the corresponding spin subsystems and not on the experimental setup details.

Integral intensity of the broad EPR absorption was determined by fitting full absorption spectra from $-5$~T to $+5$~T with the sum of two Lorentzian lines and baseline drift \eqref{eqn:p-trans}. The free paramagnetic centers absorption intervals were excluded from the fit, cavity background was taken into account if necessary.  Lorentzian lines parameters for the fit procedure were constrained as follows:  symmetric positions, same linewidths, different amplitudes and, if necessary, different dispersions. The integral intensity  of the free paramagnetic centers absorption was determined  by numerical integration using the fit of the broad resonance component as a base-line. Large difference of the spectral component linewidths allows us to resolve these components reliably, fit procedure was found to converge regularly. Intensities ratios were found
 separately for positive and negative fields allowing for possible drifts of the experiment conditions during the field scan.

(III) As a final step of the calibration protocol we calculate contribution of the EPR-active spin subsystem of the nabokoite to the total static susceptibility. We assume static susceptibility of the free paramagnetic centers to follow Curie law. The  molar susceptibility of the EPR-active spins $\chi_\textrm{EPR}$ can be now found \emph{without any additional assumptions}:

\begin{equation}\label{eqn:chi-EPR}
\chi_\textrm{EPR}=x N_\textrm{A} \frac{g^2 \mu_\textrm{B}^2}{4k_\textrm{B} T} \frac{I_\textrm{nabok}}{I_\textrm{free}}
\end{equation}

\noindent here $x$ is the concentration of the free paramagnetic centers per copper spin (see Table~\ref{tab:intdef}), mean $g$-factor value was taken to be 2.2. The susceptibility of the EPR-active spins is calculated independently for the positive- and negative-field absorption signals, found values $\chi_\textrm{EPR}^{(+)}$ and $\chi_\textrm{EPR}^{(-)}$ were usually in good agreement with each other indicating reliability of the used protocol. We take mean value $(\chi_\textrm{EPR}^{(+)}+\chi_\textrm{EPR}^{(-)})/2$ as a final estimate of the EPR-active spins contribution and spread of the  $\chi_\textrm{EPR}^{(+)}$ and $\chi_\textrm{EPR}^{(-)}$ as a characteristic error-bar.

 As a result, we  determine static susceptibility contribution of the EPR-active spins  with the accuracy of 10-20\%. Introduced calibration protocol  eliminates most of the error sources related to irreproducibility of experiment conditions.

\subsection{Contribution of the  EPR-active subsystem to the total magnetic susceptibility of nabokoite}

\begin{figure*}[th]
\centering
 \includegraphics[width=\textwidth]{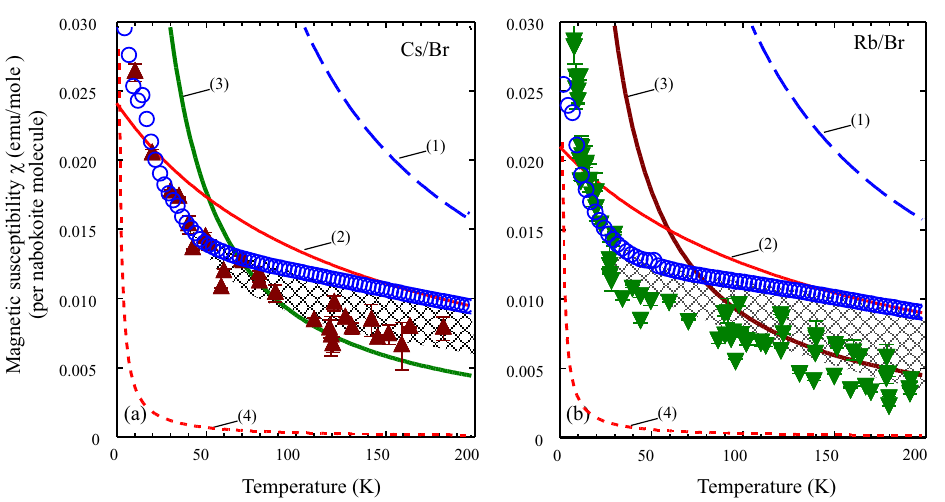}\\
  \caption{(color online) Comparison of the static susceptibility $\chi_\textrm{static}$ (open circles) and recovered contribution of the EPR-active spins to the total static susceptibility $\chi_\textrm{EPR}$ (filled triangles). Panel (a) --- data for Cs/Br nabokoite, panel (b) --- data for Rb/Br nabokoite. Curves at both panels: (1) --- Curie law for 7 spins $S=1/2$, $g=2.2$ per nabokoite molecule;  (2) ---  Curie-Weiss law for 7 spins $S=1/2$, $g=2.2$ per nabokoite molecule and Curie-Weiss  temperatures $\Theta=130$~K for Cs/Br nabokoite and 150~K for Rb/Br nabokoite;  (3) ---  Curie law for 2 spins $S=1/2$, $g=2.2$ per nabokoite molecule;  (4) --- contribution of free paramagnetic centers (see Table~\ref{tab:intdef}). Hatched areas  highlight deficient part of the susceptibility $\Delta\chi=\chi_\textrm{static}-\chi_\textrm{EPR}$.}\label{fig:suscept}
\end{figure*}

\begin{figure}[th]
\centering
  \includegraphics[width=\columnwidth]{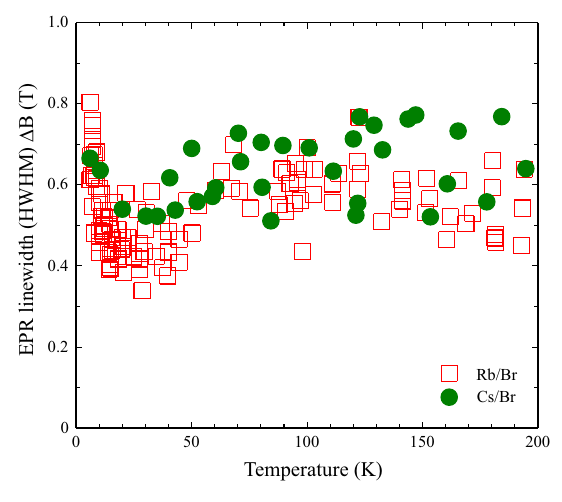}\\
  \caption{Temperature dependence of EPR line width (half-width at half-maximum) for Cs/Br (filled circles) and Rb/Br (open squares) nabokoites. $f=38$~GHz.}\label{fig:linewidth}
\end{figure}

For quantitative analysis of the absolute intensity of EPR absorption we choose two of the nabokoite family members: \CsBr{} and \RbBr{}. For both samples we have followed calibration protocol described in previous subsection and determined magnetic susceptibility of the spin subsystem responsible for the observed EPR absorption (EPR-active subsystem). Resulting data are shown at Fig.~\ref{fig:suscept}.  Results obtained for both field polarities follow the same trend with noise due to some unaccounted experimental issues. This noise is about 20\% which is above the uncertainties of the calibration protocol. Additionally this analysis provides temperature dependence of the EPR linewidth  (Fig.~\ref{fig:linewidth}).

The determined contribution of the EPR-active subsystem to the samples susceptibility can be compared with the static susceptibility (Fig.~\ref{fig:suscept}). Note that the static susceptibility data were not used in our calibration protocol.  Susceptibilities of the EPR-active subsystems both for \CsBr{} and \RbBr{} behave similarly compared to the total static susceptibility of the samples. At low temperatures (10-30~K) determined susceptibility of the EPR-active subsystem approaches static susceptibility curve within the accuracy margin of our calibration protocol. I.e., at low temperatures we observe EPR signal from all spins, contributing to static magnetization of nabokoites. However, at higher temperatures (100-200~K) susceptibility of the EPR-active subsystem amounts to just a fraction of the total susceptibility.
As a tentative estimate of the amount of the EPR-active spins we note that their contribution to the total spin susceptibility above 100~K is close to two free paramagnetic spins per nabokoite molecule (see curves (3) at Fig.~\ref{fig:suscept}).

These observations directly point that \emph{at high temperatures} fraction of magnetically active (i.e., contributing to the total static susceptibility)  spins does not contribute to the observed EPR signal, while \emph{at low temperatures} this ``EPR-silent'' fraction vanishes. This means that the spin system of nabokoite is split into decoupled spin subsystems.   Since the observed EPR signal transforms into antiferromagnetic resonance signal at N\'{e}el point without sharp changes in intensity, we conclude that magnetic order is formed within the EPR-active subsystem. Difference between the total static susceptibility and the recovered contribution of the EPR-active subsystem $\Delta\chi=\chi_\textrm{static}-\chi_\textrm{EPR}$ decreases with cooling. Experimental uncertainties became quite large after subtraction of close quantities at low temperatures, but without additional assumptions we can claim that the ``missing'' part of susceptibility $\Delta\chi$ decreases at least two-fold on cooling from 200~K to 15~K. This could indicate gapped excitations spectrum in the ``EPR-silent'' subsystem. The gap can be only roughly estimated as $\Delta \simeq 50-100$~K from the characteristic temperature of the crossover from high-temperature to low-temperature regimes, experimental uncertainties prevent reliable thermoactivation law analysis of deficient susceptibility part $\Delta\chi$.

Same analysis yields temperature dependences of EPR linewidth (Fig.~\ref{fig:linewidth}). Both for \nabok{Rb}{Br}{} and \nabok{Cs}{Br}{} linewidth is almost temperature independent from 50 to 200~K, its value is about 0.6~T. There is a tendency of slight line broadening below 20~K on approaching the N\'{e}el temperature.

\section{Discussion}

Main experimental observation of this research relies on calibration procedure that allows us to eliminate most of the  experimental uncertainties. We took advantage of the presence of free paramagnetic centers (most likely, surface or volume defects of powder particles) with distinct EPR line-shape and performed two-step calibration procedure. At first step concentration of free paramagnetic centers was determined by comparison with reference sample. At second step free paramagnetic centers themselves serve as a reference to quantitatively determine amount of spins in the sample responsible for the observed EPR absorption. This allowed us to separate model-independently contribution of the EPR-active spins to the static susceptibility of the nabokoite samples at broad temperature range from the N\'{e}el point to approx. 200~K (Fig.~\ref{fig:suscept}), which turns out to be just a fraction of  the total static susceptibility. The later observation directly implies that the spin system of nabokoite is split into decoupled spin subsystems: EPR-active subsystem which orders at the N\'{e}el point and EPR-silent subsystem. Contribution of the EPR-silent subsystem to the total static susceptibility of nabokoites vanishes on cooling.

The fact that magnetic (contributing to the total static susceptibility) subsystem remains EPR-silent can be explained by the high relaxation rate $1/\tau$ within this spin subsystem and, consequently, very high EPR linewidth $\Delta B \simeq \hbar/(\tau \mu_\textrm{B})$. Note, that susceptibility is proportional to the integral intensity of the EPR absorption, and the later is proportional to the product of the linewidth and absorption amplitude. I.e., increase of the linewidth yields proportional loss in the absorption  amplitude.  Absorption signal of the EPR-active subsystem has a linewidth about 0.6~T and is already quite tricky to observe. If the relaxation rates in  EPR-silent system would be, e.g.,  fivefold as high --- this absorption signal would be too broad and too weak to be reliably resolved in our setup.

Such an assumption of the large  EPR linewidth is not unheard of.  EPR linewidth in concentrated magnets is usually determined by anisotropic spin-spin interactions and spin-lattice relaxation processes. At temperatures about the liquid nitrogen temperatures (which are below typical Debye temperatures) spin-spin relaxation usually dominates. This is in agreement with the case of nabokoites as linewidth of the observed EPR absorption is almost temperature independent at the temperature range where phonon population numbers change strongly. Local symmetry of all important exchange bonds in nabokoites is low enough to allow for Dzyaloshinskii-Moriya (DM) interaction. The Dzyaloshinskii-Moriya interaction arises as a first-order spin-orbital coupling correction to the Heisenberg exchange, the same spin-orbit coupling  provides $g$-factor anisotropy $\Delta g$, which yields conventional estimate \cite{a-blean} of DM interaction strength $D\simeq (\Delta g/g) J \simeq 0.1 J$ since typical $g$-factor anisotropy in cuprates is about 10\%. DFT   calculations combined with symmetry analysis \cite{streltsov} demonstrated that in the related compound with decorated SKL spin system Na$_6$Cu$_7$BiO$_4$(PO$_4$)$_4$Cl$_3$ strength of DM interaction can reach up to 28~K in agreement with the estimations above.

Strong Heisenberg exchange coupling additionally introduces exchange narrowing mechanism \cite{altkoz,a-blean}, which yields for the linewidth estimate

\begin{equation}\label{eqn:dBestimate}
\Delta B\simeq g \mu_\textrm{B} \frac{(\Delta B_0)^2}{J}\simeq \frac{1}{g\mu_\textrm{B}} \frac{D^2}{J}\simeq 0.01 \frac{J}{g\mu_\textrm{B}}
\end{equation}

\noindent This rough estimate ignores orientational dependence of the linewidth, number of important bonds and other fine details which could result in an unknown factor. Estimating characteristic exchange coupling to be about the Curie-Weiss temperature value,  $J=150$~K, one obtains $\Delta B\simeq 1.0$~T. This estimation demonstrates, that it is quite plausible that the observed linewidth of the EPR-active subsystem can be ascribed to DM interaction. Supposed multiple increase of the linewidth in EPR-silent subsystem can be ascribed to unaccounted factors in rough estimate \eqref{eqn:dBestimate}.

EPR data alone cannot provide information on the microscopic origin of the found separation of spin subsystems in nabokoites. Some local probes (e.g. NMR on copper nuclei) or determination of the magnetic structure with neutron scattering experiments probably can visualize these spin subsystems. The chlorine NMR data of \cite{china} are probably not sensitive to the effects of spin subsystems separation as there is single Cl-ion crystallographic position in nabokoites (see Fig.~\ref{fig:skl-and-struct}), therefore chlorine nucleus NMR response somehow average contributions of copper ions in different positions.

Our experiment reveals, that contribution of the EPR-active subsystem amounts to about 1/2 of the total spin susceptibility of the nabokoite. Quantitative estimation of the amount of spins in the EPR-active subsystem is a subject of quite large error margins depending on model assumptions and taking into account experimental uncertainties. As shown at Fig.~\ref{fig:suscept} the recovered contribution of the EPR-active subsystem to the total susceptibility of the nabokoite sample is close to the contribution of two paramagnetic spins per nabokoite molecule. Keeping in mind experimental uncertainties, lower estimate of one spin per nabokoite molecule cannot be totally excluded. On the other hand, antiferromagnetic ordering within the EPR-active subsystem implies relevant antiferromagnetic couplings and, hence, a Curie-Weiss law with antiferromagnetic Curie-Weiss temperature, which suppresses  magnetization and raises estimation of the spins fraction involved. Thus, we can reasonably conclude that the EPR-active subsystem includes from one to three spins per nabokoite molecule.

Structure of nabokoites (Fig.~\ref{fig:skl-and-struct}) and theoretical analysis of exchange bonds strengths in K/Cl nabokoite \cite{gonzales} provide two speculative possibilities for the observed subsystems splitting.

Firstly, we can consider structure of the decorated square kagom\'{e} lattice (Fig.~\ref{fig:skl-and-struct}) which includes decorating interlayer copper ion in Cu2 position which is a natural candidate for the role of decoupled spin. It is bridged to the SKL layer with four equivalent bonds, which compensate each other within the mean field approximation. I.e., Cu2 spins, one spin per nabokoite molecule, are the candidates for the EPR-active subsystem in this model. SKL layers are predicted to be a gapped spin-liquid  for the equilateral lattice \cite{Richter-skl-gapped}, the gapped spin-liquid state should persist in a certain parameters range in the case of nonequivalent exchange bonds. The SKL layers with gapped excitations spectrum are then the candidates for the EPR-silent subsystem. Note, that in this case ordering of the EPR-active subsystem of Cu2 spins requires unusual RKKY-like coupling via excitations in SKL layers, as was proposed in \cite{markina}.

Secondly, we can consider structure of exchange bonds proposed for K/Cl nabokoite by DFT calculations \cite{gonzales}. These DFT calculations predict strong NNN coupling Cu3-Cu3 ($J_6$ in terms of \cite{gonzales}) which amounts to approx. 20\% of the strongest exchange coupling in nabokoite. This allows us to consider spin system of nabokoite as simple square lattice formed by Cu3 ions which is decorated by a checkerboard pattern of Cu1-Cu2 pyramids. Within this model Cu3 ions, which are two  per nabokoite molecule, are candidates for the EPR-active subsystem, and five-spins Cu1-Cu2 pyramids are the candidates for the EPR-silent subsystem. As pyramid bridging to the Cu3 layers is frustrated, they can be effectively decoupled from the Cu3 layers at certain coupling parameters. Antiferromagnetic ordering within the simple square lattice of Cu3 ions requires only weak interlayer coupling or Ising anisotropy. An odd number of spins in Cu1-Cu2 pyramids mean that they should be magnetic at low temperatures, however there is a possibility of thermal crossover from the low temperature collective $S=1/2$ state of the five-spins cluster to the high-temperature limit of five independent spins $S=1/2$ which could explain decrease of the EPR-silent subsystem contribution to the susceptibility on cooling.

The simple  model of fully separated spin subsystems can be too naive for the real compounds. E.g., Monte-Carlo spin structure factor analysis for the proposed exchange bonds geometry \cite{gonzales} demonstrates that all copper ions positions (Cu1, Cu2 and Cu3) develop regular spin correlations patterns at low temperatures. However, the simple considerations above demonstrate that the exchange bonds geometry of nabokoite compounds includes motives leading to possible separation of spin subsystems in agreement with the experimental observations.

\section{Conclusions}

We report results of electron paramagnetic resonance study of the highly frustrated nabokoite family magnets \nabok{A}{X}{} (A=Na, K, Rb, Cs; X=Cl, Br). We have found broad (HWHM linewidth 0.3...0.7~T) paramagnetic resonance absorption, the observed linewidth is qualitatively in agreement with the presence of Dzyaloshinskii-Moriya interaction. 

We have performed accurate calibration of the EPR absorption which revealed that EPR-active spin subsystem of nabokoite contributes approx. 50\% of the total static spin susceptibility at high temperatures. Remaining susceptibility arises from the EPR-silent subsystem, which does not contribute remarkably to the observed EPR absorption due to the high relaxation rate within this subsystem.

Thus, electron paramagnetic resonance experiments reveal the  existence of the decoupled spin subsystems within the complicated geometry of the exchange bonds in nabokoite.

\acknowledgements

Authors thank Prof.~A.~Smirnov, Prof.~L.~Svistov and Dr.~S.~Sosin (Kapitza Institute) for stimulating discussions.

The work at P.~Kapitza Institute for Physical Problems (magnetic resonance experiments and data analysis) was supported by RSF~22-12-00259-$\Pi$.   Some of the authors (MMM and ANV) acknowledges the support by the Ministry of Science and Higher Education of the Russian Federation in the framework of the Strategic Academic Leadership program ``Priority 2030'' (MISIS Strategic Technology Project `Quantum Internet').

\end{document}